\documentclass[aps,prb,twocolumn,groupedaddress]{revtex4}
\usepackage[dvips]{graphicx,color}
\usepackage{bm}
\newcommand{\beq}{\begin{equation}}
\newcommand{\eeq}{\end{equation}}
\newcommand{\beqa}{\begin{eqnarray}}
\newcommand{\eeqa}{\end{eqnarray}}
\newcommand{\figl}{\includegraphics[scale=1,draft=false,width=3.2in,height=3.3in,keepaspectratio=true]}
\newcommand{\figs}{\includegraphics[scale=10,draft=false,width=6.5in,height=5.5in,keepaspectratio=true]}
\newcommand{\befig}{\begin{figure}[h]}
\newcommand{\efig}{\end{figure}}

\begin{document}

\title{Correlated quantum percolation in the lowest Landau level}

\author{Nancy Sandler}
\email{sandler@ohio.edu}
\affiliation{Dept. of Physics and Astronomy, Ohio University, Ohio, OH
  45701, USA}
\author{Hamid R. Maei}
\email{maei@brandeis.edu}
\affiliation{Gatsby Unit, University College London, London, UK}
\author{Jan\'{e} Kondev}
\email{kondev@brandeis.edu}
\affiliation{Department of Physics, Brandeis University ,Waltham, MA
  02454, USA }

\date{\today}

\begin{abstract}
Our understanding of localization in the integer quantum Hall effect
is informed by a combination of semi-classical models and percolation
theory. Motivated by the effect of correlations on classical
percolation we study numerically electron localization in the lowest
Landau level in the presence of a power-law correlated disorder
potential. Careful comparisons between classical and quantum dynamics
suggest that the extended Harris criterion is applicable in the
quantum case. This leads to a prediction of new localization quantum
critical points in integer quantum Hall systems with power-law
correlated disorder potentials. We demonstrate the stability of these
critical points to addition of competing short-range disorder
potentials, and discuss possible experimental realizations.
\end{abstract}
\maketitle

\section{Introduction}
\label{sec:intro}

One of the hallmarks of the quantum Hall effect is the appearance of
plateaus in the Hall resistance of a two-dimensional electron gas as a
function of applied magnetic field. Transitions between plateaus have
been associated with a quantum critical point \cite{ref:Pruisken}.
This critical point corresponds to a localization-delocalization
transition for the electron's wave-function. Unlike the Anderson
metal-insulator transition here the transition is between two
insulating states: an extended wave function emerges at a single value
of the magnetic field corresponding to the center of a Landau
band. Experiments and numerical simulations have revealed many
important properties of this critical point \cite{ref:Huckestein1},
such as the value of the correlation length exponent, $\nu_q \approx
2.3$.  Attempts at formulating an analytically tractable theory of the
transition have by and large failed.

A very fruitful approach to unraveling the physics of plateau
transitions has been the semi-classical description of electron
dynamics in the lowest Landau level \cite{ref:Trugman}. Classically
the electron's motion can be described as a fast cyclotron rotation in
the plane accompanied by a slow ${\bf E}\times {\bf B}$ drift of the
center of the circular orbit along lines of constant random potential,
$V({\bf x})=$const.  Quantum mechanics adds tunneling and
interference to the mix.

The purpose of this work is to understand the relation between
classical and quantum dynamics of electrons in the lowest Landau
level, in the presence of a random potential. In particular we
investigate the effect of the fractal geometry of the classical orbits
on the critical properties of the quantum localization-delocalization
transition, such as the value of $\nu_q$.  One of our main results is
that we have identified a family of random potentials for which the
classical electron orbits have continuously varying fractal
properties, while the quantum critical point remains unchanged. These
random potentials are characterized by decaying power-law correlations
in space. When the decay of correlations is not too fast we find new
quantum critical points which are the quantum analogs of correlated
percolation \cite{ref:Weinrib2}.

Previously, a real-space renormalization group treatment of correlated
quantum  percolation appeared in Ref.~\onlinecite{ref:Cain}.
Preliminary results of our numerical approach were reported in
Ref.~\onlinecite{ref:short}. Here we provide a detailed
account of the numerical methods used to study the effect of power-law
correlated disorder potentials on the integer quantum Hall transition
(IQHT),  as well as theoretical arguments supporting them. We also
describe new results on the stability of these quantum critical points
to the presence of short-range correlated disorder.
The stability analysis provides connections with experiments
\cite{ref:Zielinski} where these new quantum critical points might be
observed.

The work presented here addresses two different aspects of the
problem. On one side the motivation is provided by understanding the
role of disorder in determining the transport properties of quantum
Hall systems. On the other side, we are motivated by the possibility of
extending correlated percolation results in the classical regime to
the quantum regime.

\subsection{Disorder and the quantum Hall effect}
\label{sec:localization}

The role of disorder in the quantum Hall effect was recognized shortly
after its discovery. From scaling theory it is known that electron
wave-functions are localized by potential disorder in two-dimensions
\cite{ref:gang4}. This remains true in the presence of a strong
perpendicular magnetic field for states in the tails of the Landau
bands, and leads to the observed plateaus in the Hall resistance
\cite{ref:laughlin}.  Toward the center of a Landau band the
localization length increases signaling the appearance of a
delocalized state. This results in the observed transition between
plateaus in the Hall resistance and the peak in the longitudinal
resistance.

The precise nature of the disordered potential experienced by
electrons in quantum Hall systems remains poorly understood. The
canonical picture (say, for GaAs heterostructures) is that ionized
donors located between the two-dimensional electron gas and the sample
surface are responsible for the random potential. Important
experimental progress has come from scanning techniques which provide
a map of the electrostatic potential seen by electrons in a
semiconductor heterostructure \cite{ref:shayegan}. Interestingly
enough the experiment measures random potential fluctuations that
extend over distances that are more than an order of magnitude larger
than the typical spacing between the impurities and the electron gas,
suggesting that the potential might be longer ranged that previously
thought. Scanning techniques and the ability to manipulate single
atoms on surfaces also opens up the possibility of engineering the
random potential so as to produce novel transport effects. For
example, an enhancement of the conductivity (compared to the Drude
value) of a two-dimensional electron gas was observed in
Ref.~\onlinecite{ref:Zielinski} due to the presence a random magnetic
field. In this case the random field was produced by a thin film of
rough magnetic material brought in close proximity to the electron
gas.  The fluctuations of the field were inherited from the height
fluctuations of the films surface.

Inspired by these experimental advances here we describe how power-law
correlated disorder can lead to novel critical behavior of the
electron gas in the integer quantum Hall setting. We also analyze the
effect of competing disorders, which are necessarily present in semiconductor
heterostructures, on the stability of these critical points.

\subsection{Quantum percolation}
\label{sec:quantumperco}

The classical percolation problem provides one of the most intuitive
examples of a critical point. Lattice sites are occupied with
some probability $p$.  Nearest neighboring sites which are occupied form
clusters with a typical size $\xi$.  As $p$ is tuned to its critical
value $p_c$, which depends on the lattice type, the correlation length
$\xi$ diverges as $\xi \sim |p_c - p|^{-\nu_c}$.  $\nu_c$ is the
correlation length exponent, and in two dimensions $\nu_c=4/3$ independent of the
lattice type \cite{ref:Dennijs}.

An analogous picture has been put forward to describe the localization
transition in the integer quantum Hall system \cite{ref:Trugman}. For
a smooth random potential (one that varies little on the scale of the
magnetic length) the electron wave-function is localized along the
level lines.  For random potential symmetric around $V=0$ level lines
away from the zero level are closed and have a typical size $\xi$
which is identified with the localization length. As the electron's
energy is tuned to the center of the Landau band the corresponding
level line approaches $V=0$ and $\xi$ diverges. If the random
potential has short range correlations in space this is precisely the
classical percolation problem \cite{ref:Isichenko} and one would predict a
localization length exponent $\nu_q=\nu_c=4/3$.

Experiments that measure a localization length exponent in the integer
quantum Hall system find $\nu_q \approx 2.3$, signaling a break-down
of the simple percolation picture. The essential physics that has been
left out is one of electron tunneling, which readily occurs at the
saddle points of the random potential, as well as quantum
interference. This was beautifully demonstrated by Chalker and
Coddington who proposed a lattice model which takes into account these
purely quantum effects, by describing electron dynamics as hopping
from one saddle point to the next with scattering matrices associated
with each saddle \cite{ref:Chalker1}. The saddle points themselves occupy the
vertexes of a square
lattice. Computer measurements of the localization exponent for this
model yield values in agreement with the experimental
results. Moreover, taking a classical limit of the network model, in
which the scattering matrices become classical probabilities for the
two possible outcomes of a scattering event at a vertex, leads to
classical percolation and a localization length exponent of
$4/3$ \cite{ref:DHLee}.  This provides an intriguing connection
between the critical points of classical and quantum percolation. Here
we investigate the nature of this connection when power-law
correlations are introduced in the random potential. In the classical
case it was shown that this can lead to new critical points if the
power-law decay is not too fast. We show that a similar effect occurs
for the quantum version of percolation provided by electron
localization in the lowest Landau level.

The paper is organized as follows. In section~\ref{sec:LLLmotion} we
discuss the physics of classical and quantum electron motion in a
disordered potential in the presence of a strong magnetic field. We
present scaling arguments that show how the value of the critical
exponent $\nu$ can be extracted from the time dependent  mean
square displacement (classical motion) and the time dependent
wave-packet spread (quantum motion). In section~\ref{sec:shortrange} we
describe in detail the numerical methods used to obtain the value of
$\nu$ in the classical and quantum setting.  As evidence of the
suitability of these methods, we present results for short-range
correlated disorder potentials that are in good quantitative agreement
with previous experimental and theoretical works
\cite{ref:Chalker1,ref:Huckestein1,ref:DHLee}.
Section~\ref{sec:longrange} contains the numerical results obtained
for classical and quantum electron motion in power-law correlated
disorder potentials. In section~\ref{sec:HarrisCrit} we propose a
framework based on the classical Harris criterion \cite{ref:Harris}
and its extension \cite{ref:Weinrib1} that provides the theoretical
support for our numerical results.

Finally, in section~\ref{sec:short-long-correl} we present results on
classical and quantum localization in the presence of competing
short-range and power-law correlated disorders. In experiments
\cite{ref:Zielinski}, which we believe have a good chance of observing
the newly discovered quantum critical points, both types of disordered
potentials are present. The key question here is whether correlated
quantum percolation is stable to the addition of short-range
correlated disorder.

\section{Classical and quantum motion in the lowest Landau level}
\label{sec:LLLmotion}

The methods and tools used to describe the localization-delocalization
transition in the lowest Landau level (LLL) in the classical regime
are considerably different from the ones used in the quantum
regime. In this section, we review both approaches and present the
arguments leading to the connection between the localization exponent
$\nu$ and electron dynamics.

\subsection{Classical motion}
\label{sec:classical-motion}

The two-dimensional semi-classical motion of an electron in a disorder
potential under a strong constant magnetic field (along the
$\hat{z}$-axis), is described by the drift motion of the electron's
guiding center along equipotential lines:

\beq
\frac{d{\bf r}}{dt}\,=\,\frac{c}{eB}\,{\nabla V({\bf r})}\,\times\, \hat{z};
\label{eq:classeq}
\eeq

where ${\bf r}$ is the position vector for the guiding center in the
$(\hat{x},\hat{y})$ plane.  This equation is derived in the adiabatic
approximation, {\it i.e.}, under the assumption that the cyclotron
radius $\ell_c=v/\omega_{c}$ is much smaller than the typical distance
over which the potential changes \cite{ref:Northrop}. Therefore the
localization-delocalization transition in the LLL in the classical
regime, implies the study and characterization of the
trajectories determined by this equation.

Eq.~(\ref{eq:classeq}) was extensively studied in two different albeit
related contexts. F.~Evers \cite{ref:Evers} used this equation in
numerical studies of classical motion along the hulls of percolation
clusters. He demonstrated interesting scaling behavior of the
time-dependent density-density correlation function close to the
percolation critical point.

V.~Gurarie {\em et al.}\cite{ref:Gurarie} arrived at
Eq.~(\ref{eq:classeq}) from the classical limit of the Liouvillian
equation of motion for electrons in the lowest Landau level of the
integer quantum Hall regime. In both cases, the authors considered
smooth random potentials with short-range correlations in space.

One goal of this work is to study numerically properties of closed
trajectories determined by Eq.~(\ref{eq:classeq}), where $V({\bf r})$
is power-law correlated in space. In particular, we focus on the
scaling law, and its associated critical exponent $\nu$, which
describes how the size of a closed trajectory grows as the energy
approaches the critical value determined by the $V({\bf r})=0$
equipotential.

To make explicit the connection between $\nu$ and the dynamics of a
particle moving in a two-dimensional random potential, let us briefly
review the argument presented in Ref.~\onlinecite{ref:Gurarie}.  For a
fixed constant value of $V({\bf r}) = V_0$, there is a set of
equipotentials associated with closed electron trajectories. These
trajectories are the outer boundaries (hulls) of percolation clusters,
with the occupation probability $p$ of the percolation problem
determined by $V_0$. In particular, the critical value of $p_c = 1/2$
corresponds to $V_0 = 0$ in the units chosen. Several studies
\cite{ref:Gurarie,ref:Evers,ref:Ziff2} have shown that the mean
squared displacement between two points on a given trajectory $\Delta
\mathbf{r}^2$ ($\mathbf{r}$ is the position vector for the particle on
the trajectory), when averaged over all trajectories for fixed $V_0$,
follows a diffusive law:

\beq
\langle \Delta \mathbf{r}^2(t) \rangle_{V_0}\, = \, D \, t ~~~~ (t \,\ll\, t^*) .
\label{eq:short}
\eeq

Here $t^*$ is a characteristic time that depends on $V_0$.

Away from the percolation critical point particle trajectories are
closed, and in the long-time limit the mean squared displacement
reaches a constant value

\beq
\langle \Delta \mathbf{r}^2(t)\rangle_{V_0}\, = \, \xi^{2} ~~~~ (t \, \gg\,  t^*),
\label{eq:long}
\end{equation}
where $\xi(V_0)$ is the localization length.

These considerations lead to a scaling form (at fixed
$V_0$) for the mean squared displacement

\beq
\langle \Delta \mathbf{r}^2(t)\rangle_{V_0}\,=\,D\,t\,f\left(\frac{D\,t}{\xi^2}\right) \ ,
\eeq
where $f$ is a scaling function.  The classical result from
percolation theory\cite{ref:Ziff2} for the correlation length, $\xi
\propto |p-p_c|^{-\nu_{c}}$, leads to the expression $\xi \propto
V_0^{-\nu_{c}}$, which in turn implies

\beq
\langle \Delta \mathbf{r}^{2}(t)\rangle_{V_0} \, =\, D\, t \,f\left(\frac{D\,t}{V_0^{-2\nu_{c}}}\right).
\label{eq:scaling}
\eeq

Finally, by averaging this equation over all values of $V_0$ one
obtains the scaling relation

\beq
\overline{\langle \Delta \mathbf{r}^2(t)\rangle}\,\,\sim\,\, t^{\theta},
\eeq
where $\theta=1-1/2\nu_{c}$ is the anomalous diffusion exponent. From
simulations  $\overline{\langle
\Delta \mathbf{r}^2(t)\rangle}$ can be computed which leads to a value for
$\theta$ and, via the scaling relation just derived, to a value for
the localization length exponent $\nu_{c}$ in the classical regime.

\subsection{Quantum motion}
\label{sec:quantum-motion}

One approach to study the localization-delocalization transition in
disordered quantum systems is based on the different contributions
that extended and localized electron wave-functions make to the
frequency dependent electrical conductivity
\cite{ref:Thouless1}. Localized states appear in the electrical
conductivity through the retarded density-density Green's function, a
quantity that has an intuitive interpretation when written in terms of
wave-functions. When the electron is represented by a localized
wave-packet at position ${\bf r}$ and time $t=0$, the Green's function
gives the value of the wave-packet spread at time $t$. Since
this wave-packet describes the probability of finding the electron at a
given position and time, its spread is a measure of the uncertainty on
the electron's position, and as such, an indicator of
localization.

In most studies of localization, the Green's function is calculated in the
momentum-frequency domain and studied in detail in the limit of large
momenta/small frequency \cite{ref:Chalker2,ref:Huckestein1}. The purpose of
this section is to show explicitly how the
critical exponent $\nu_q$ can be extracted from the Green's function
using instead the real time domain, as proposed by Sinova, Meden and
Girvin \cite{ref:Sinova}.

Within this approach, one studies the disorder averaged
density-density correlation function projected onto the LLL, i.e.,
the focus is on the unconstrained spectral function.  As
pointed out in Ref.\onlinecite{ref:Gurarie}, the applicability of the
method rests upon the assumption that delocalized states in the
IQH transition are isolated from each other and located at
the centers of the Landau bands. Thus, when the density-density
correlation function is restricted to one Landau
level (the lowest being considered for simplicity), the contribution
of the (single) delocalized state in that level dominates the sum over
states in the small $(q, \omega)$ limit. Equivalently, the integral of
the spectral function over all energies is dominated by the
contribution from the critical energy in that limit.
For our purposes,
the main advantage of this approach is that it clearly spells out a
numerical scheme for studying the effect of power-law correlated disorder
potentials on localization in the lowest Landau level.

In order to make connections with electron motion in the classical regime
and also to fix the notation used in the rest of the paper, we
review some of the main points of the approach. Consider 2d spinless
electrons (the electron spin is fixed by the magnetic field and we
assume that spin-flipping interactions are not allowed) in the x-y
plane, under the combined effects of a magnetic field ${\bf B} = B
\hat{z}$, perpendicular to the plane and a random potential
$V({\bf r})$ due to the presence of impurities.

The electron-impurity interaction is given by:

\beq
H\, =\sum_{{\bf k}} V_{-{\bf k}}\, \rho_{{\bf k}} \ ,
\eeq
where $\rho_{{\bf k}} = e^{i{\bf k}\cdot{\bf r}}$ is the one-particle
density operator and $V_{{\bf k}}$ is the Fourier transform of the
disorder potential.  At high enough magnetic fields (or low enough
temperatures), the kinetic energy is quenched to the value of the
lowest Landau band and the sum is reduced to a sum over states in the
LLL. Let $\bar{H}$ and $\bar{\rho}$ be the Hamiltonian and
one-particle density operator projected onto the LLL. Then,

\beq
\bar{H}\, =\, \sum_{{\bf k}}\, V_{-{\bf k}}\,\bar{\rho}_{{\bf k}},
\label{eq:hamilt}
\eeq
with the projected one-particle density operator taking the form:
\beq
\bar{\rho}_{{\bf k}}= e^{-\ell^2_c\,k^2/4}\,e^{\imath {\bf k}\cdot{\bf C}}
\label{eq:denproj}
\eeq
with ${\bf C}$, the position vector for the guiding center of a
one-particle orbit, defined as \cite{ref:Ezawa}:
\beqa
C_x\,&=&x\,-\,\frac{c}{eB}\,\Pi_y \nonumber \\
C_y\,&=&y\,+\,\frac{c}{eB}\,\Pi_x \ .
\label{eq:guidecent}
\eeqa

Here ${\bf \Pi}={\bf p}+\frac{e}{c}{\bf A}$ and $(x,y)$ are the
canonical momentum and position operators for one particle.

The formalism used to project the density operator
$\rho_{{\bf q}}\,=\,e^{-i{\bf q}\cdot{\bf r}}$ onto the LLL was
developed in Refs.~\onlinecite{ref:Girvin1,ref:Girvin3}. There it was
shown that the commutation relation for the projected density operator
obeys a closed algebra (the magnetic translation operator algebra):

\beq
[ \bar{\rho_{{\bf k}}}\,,\,\bar{\rho_{{\bf q}}} ]\,=\, 2\;i\; \sin\left(\frac{\ell_c^2}{2}k \times q\right)\;e^{\frac{-\ell_c^2}{2}{\bf q}\cdot{\bf k}}\; \rho_{{\bf k}+{\bf q}},
\label{eq:commut}
\eeq

with $k \times q =({\bf k}\times {\bf q})\cdot\hat{z}$. This property
implies that the equation of motion for the density operator

\beq
\imath\, \hbar \,\frac{\partial}{\partial t}\bar{\rho}\, =\, [\bar{H} , \bar{\rho}]
\eeq
is closed and can, in principle, be solved exactly.

Following the same line of argument, one can show that similar results
hold also for the correlation function of the projected density
operator:

\beq
G({\bf r},t)\, =\, Tr\left(\bar{\rho}({\bf r},t)\, \bar{\rho}({\bf 0},0)\right)
\eeq
note that the trace extends over states in the lowest
Landau level only. The density-density correlation function $G$ in
${\bf k}$-space satisfies the equation \cite{ref:Sinova}:

\beqa
\imath \,\hbar\, \frac{\partial}{\partial t}G({\bf k},t)\, &=& \,
\sum_{{\bf q}}\, 2\;i\; \sin\left(\frac{\ell_c^2}{2}k \times q\right)\;V({\bf k}-{\bf q}) \nonumber \\
& & \exp[-\frac{\ell_c^2}{2}(k^2-{\bf k} \cdot {\bf q})]\, G({\bf q},t)
\label{eq:greensfunc}
\eeqa 
which follows from Eq.~(\ref{eq:commut}). This is a
Schr\"oedinger-like equation where the effective Hamiltonian is the
Liouvillian matrix:

\beq
{\cal{L}}_{{\bf k}{\bf q}} =  2\,\imath\, \sin\left(\frac{\ell_c^2}{2}k \times q\right)\,V({\bf k}-{\bf q})\,\exp[-\frac{\ell_c^2}{2}(k^2-{\bf k} \cdot {\bf q})]
\eeq

(The name Liouvillian is used because of the analogy with the
Liouvillian operator in classical mechanics.)

An interesting feature of Eq.~(\ref{eq:greensfunc}) is that for a
fixed value of the magnetic field, the magnetic length $\ell_c$
($\ell_c^{2}=\frac{\hbar c}{eB}$), vanishes in the classical limit
$\hbar \rightarrow 0 $. In units of $\frac{c}{eB}=1$,
Eq.~(\ref{eq:greensfunc}) gives\cite{ref:Gurarie}:

\begin{equation}
\frac{\partial}{\partial t}G({\bf r},t)=\epsilon_{ij}\partial_{i}V({\bf r})\partial_{j}G({\bf r},t).
\label{eq:driftC}
\end{equation}

A solution of this equation is $G(r,t)=\delta({\bf r}-{\bf r}(t))$ where

\begin{equation}
\frac{d{\bf r}}{dt} = {\nabla V(x,y)}\,\times\, \hat{z} \ .
\label{eq:classic}
\end{equation}

In other words, Eq.~(\ref{eq:driftC}) describes the classical motion
of the electron's guiding center along equipotentials of $V({\bf
r})$\cite{ref:Ezawa}.

As argued above, the Liouvillian driving the quantum dynamics of an
electron in the LLL, contains information about localization and the
value of the critical exponent $\nu_q$. The following scaling argument
provides a numerical algorithm for extracting the localization
exponent from the time dependent projected density-density correlation
function. (We would like to draw attention to a close analogy between
the present argument and the ideas presented in
Sec.~\ref{sec:classical-motion}.)

Let us start with Eq.~(\ref{eq:hamilt}), and assume that all
eigenvalues and eigenstates, localized and delocalized, are known.
Assume next that the disorder potential breaks completely the
degeneracy of the LLL, and that there is only one extended state at
energy $E=0$. We construct a wavepacket at time $t=0$ using only
localized states with energies in the $\Delta$ neighborhood $E' \neq
0$, and then we let the wave-packet evolve in time. The state-vector
corresponding to such wave-packet is:

\beq 
|\psi_{E'}(t)> = \sum_{i=E'-\Delta}^{E'+\Delta} c_{i}(t) |i> 
\eeq
where $|i>$ are eigenvectors of the Hamiltonian projected onto the LLL
(see Eq.~(\ref{eq:hamilt})).

At very short times, the spread of the wave-packet as a function of
time is ballistic (see Appendix for derivation). However, after some
crossover time and before the localization length corresponding to the
energy $E'$ is reached, the spread of the wave-packet averaged over
disorder realizations is diffusive:

\beq
\overline{<\Delta r^2(t)>_{E'}}\, = \, D \, t ~~~~ (t \,\ll\, t^*(E')) \ .
\label{eq:shortxi}
\eeq

At times much longer than $t^*\sim \xi^2$, the localization length is reached and

\beq
\overline{<\Delta r^2(t)>_{E'}}\, = \, \xi^{2}(E') ~~~~ (t \,\gg\, t^*(E')) \ .
\label{eq:longxi}
\eeq

In close analogy with the classical argument, a scaling form for the
average dispersion of the wave-packet follows:

\beq
\overline{<\Delta r^2(t)>_{E'}}\,=\,D\,t\,f'\left(\frac{D\,t}{\xi^2(E')}\right) \ ,
\label{eq:defofg}
\eeq

where $f'$ is the appropriate scaling function.  Now if a localized
wave-packet at time $t=0$ is constructed with all the eigenstates
(including the lone delocalized state at the center of the LLL), its
wave-vector can be written as:

\beq
|\psi(t)> = \sum_{E'\to -\infty}^{E'=0} |\psi_{E'}(t)>  \ .
\eeq

Using this decomposition
the dispersion of $|\psi(t)>$ can be expressed as

\beq
\overline{<\Delta r^2(t)>} \simeq \int_{E'\to
  -\infty}^{E'=0}\;dE'\;\overline{<\Delta r^2(t)>_{E'}} \; g(E') \ ,
\eeq

where $g(E')$ is the density of states; the contribution of the cross
terms can be neglected assuming that the extended state is
non-degenerate.  Specifically, it can be shown that the cross terms
are a correction to the leading behavior if there is only one extended
state at the critical energy ($E=0$). Replacing $\overline{<\Delta
r^2(t)>_{E'}}$ by Eq.~(\ref{eq:defofg}), using the fact that the
density of states is non-singular and that $\xi$ is related to $E'$ by
$\xi(E') \sim (E')^{-\nu_q}$, we obtain:

\beq
\overline{<\Delta r^2(t)>} = \int_{-\infty}^0\;dE'\;D\;t\;g(E')\;f'(\frac{Dt}{E'^{-2\nu_q}})
\eeq

Finally, an appropriate rescaling gives:

\beq
\overline{<\Delta r^2(t)>} \propto t^{\theta} \ ; \ \  \theta = 1- 1/2\nu_q ,
\label{eq:ThetaNu}
\eeq
and we conclude, as in the classical case, that the spread of the
wave-packet is sub-diffusive with an anomalous diffusion exponent
$\theta$.

Therefore, by computing $\overline{<\Delta r^2(t)>}$, the value of $\nu$
can be obtained. An alternative derivation of this result has been
proposed in Ref.~\onlinecite{ref:Oganesyan},
where the energy integrated Liouvillian propagator is analyzed in the
finite $(q, \omega)$ regime and the limits of $q\to 0$ and small $\omega $
are taken.

\section{Short range disorder potential}
\label{sec:shortrange}

We introduce next the numerical procedures used to calculate the value
of the critical exponent $\nu$.  We do this first in the case of a
short-range correlated disorder potential, where we test the
effectiveness of these numerical techniques by re-deriving previously
known results. We begin by introducing the method used for analyzing
classical localization, where the connection to percolation theory
provides a testing ground for our numerics. In subsection
\ref{sec:shortquantum} we describe how the Liouvillian approach is
implemented following the method introduced in
Ref.~\onlinecite{ref:Boldyrev} to analyze quantum localization. The
numerical values thus obtained compare favorably to experimental
results and previous numerical calculations.

\subsection{Classical motion}
\label{sec:shortclassical}

To study the classical motion of the electron's guiding center along the
equipotentials of $V({\bf r})$ we make use of a lattice model, used
previously to compute geometrical exponents for contour loops of rough
interfaces \cite{ref:JKPRE}.

Consider a square lattice ${\cal L}$ with $N$ sites and periodic
boundary conditions. Each site in the lattice has assigned to it a
random number that represents the value of the disorder potential
$V({\bf r})$ at that point in space. Its dual lattice ${\cal L^*}$ is
the set of points that describe the positions of the electron in this
potential landscape. Thus, an electron originally located at a site
${\bf r}_0$ of ${\cal L^*}$ will move along a path that joins points
on this lattice. The trajectories are contour loops of the disorder
potential. They are generated by first randomly selecting the initial
position ${\bf r}_0$ from the set of points in ${\cal L^*}$.  The
choice of the initial position of the trajectory also fixes the value
of the level $V_0$. $V_0$ is computed as the average of the values for
$V({\bf r})$ at the four vertexes (sites in ${\cal L}$) that surround
the initial position ${\bf r}_0$.  Once $V_0$ is determined, all the
lattice sites of ${\cal L}$ can be labeled by $+$ or $-$, depending
on whether they are at a potential higher or lower than $V_0$.  The
numerical algorithm updates the electron's position by selecting the
bond on the dual lattice that connects the point ${\bf r}_0$ with one
of its nearest neighbor points, with the additional property that it
crosses a $+-$ bond on the lattice ${\cal L}$. With this construction,
the contour loop is a directed walk along the bonds of the dual
lattice ${\cal L^*}$, that separates potentials lying above (below)
$V_{0}$ on the inside (outside) of the loop (see
Fig.~\ref{fig1:loops}).

\befig
\noindent
\figl{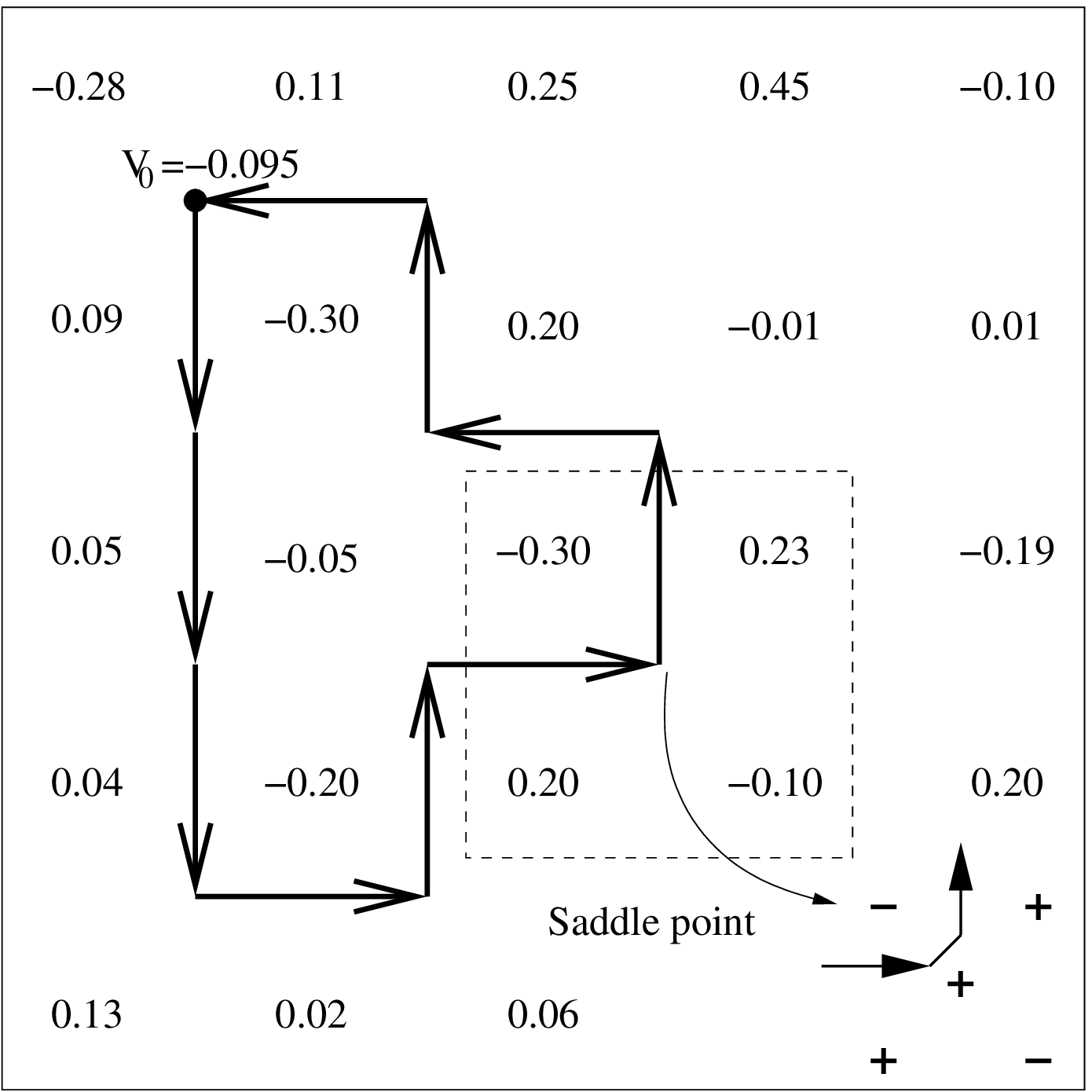}
\caption{\label{fig1:loops} Classical trajectory of the electron's
guiding center, in the presence of a random potential whose values are
indicated. $V_0$ is the average potential around the plaquette
surrounding the initial point of the trajectory. The resolution of a
saddle point is determined by the average potential around the
plaquette surrounding the saddle.}  \efig

At each step, the square of the displacement vector between the
current and original positions on the dual lattice is calculated. The
procedure is repeated for different starting points ${\bf r}_0$, and
different disorder realizations.  All the calculated quantities are
averaged over all trajectories and over all disorder realizations.

A technical subtlety of this procedure is the existence of
saddle-point plaquettes, where there are $+-+-$ signs cyclically
around them. In these cases, two contours (four links) meet at the
center and it is necessary to add an additional rule to resolve the
connectivity, so as to convert this pattern into two
$90^{\circ}$ turns that are not quite touching.  A physically sensible
rule makes use of the average potential $V_{plaq}$ of the four
potentials around the saddle-point plaquette. If $V_{plaq}<V_{0}$, the
center of the plaquette is at a lower potential than the contour loop
and the connectivity is resolved by having the $+$ sites inside the
$90^{\circ}$ turns. In the opposite case, $V_{plaq}>V_{0}$, the $+$
sites are outside of the $90^{\circ}$ turns (see
Fig.~\ref{fig1:loops}).

As in Ref.~\onlinecite{ref:Gurarie} we find from simulations that the
average time for an electron to traverse a certain distance on ${\cal
L^*}$ is proportional to the distance. This observation justifies
using the number of steps along ${\cal L^*}$ to measure the time
elapsed. In this way time is rendered dimensionless.

The computations were carried out on two-dimensional square lattices
with system sizes $N=256,512$ and $1024$. The values for the
short-range correlated potential were chosen independently from site
to site from a uniform distribution over the range $[-0.5,0.5]$. This
choice leads to $V=0$ for the value of the critical equipotential
energy. Fig.~\ref{fig2:classh} shows results for a $1024$ x $1024$
lattice obtained by averaging over $3\times 10^{4}$ trajectories and
$1000$ different disorder realizations.
\vspace{1.0cm}

\befig
\figl{Fig_2.eps}
\caption{\label{fig2:classh} Averaged mean square displacement for a
classical particle drifting along equipotentials of a random
short-range correlated potential. Distance is measured in units of
lattice spacing and time in number of lattice steps. System size is
$1024 \times 1024$.  $\nu$ is related to the slope in the critical
region $\theta$, by $\theta=1-1/2\nu$.}  
\efig

As discussed above, the diffusive (slope $\theta = 1$), short-time
($t< t^* \sim 10$), behavior is followed by a crossover to a critical
regime where the value of the slope $\theta$ is a measure of the
critical exponent $\nu_{c}$; $\theta = 1-1/2\nu_c$. By fitting the
critical region to a line on a log-log graph we find

\beq
\nu_{c}= 1.25 \pm 0.05
\eeq
where the error bars reflect the uncertainty associated with choosing
the critical region.

This value compares well with the exact result from percolation theory
$\nu_{c}= 4/3$ \cite{ref:Dennijs}.  The critical regime extends until
finite size effects become dominant and $<\Delta r^2>$
saturates, as observed in Fig.~\ref{fig2:classh} for $ t > 2 \times
10^4$.

\subsection{Quantum motion}
\label{sec:shortquantum}

Following the ideas presented in Sec.\ref{sec:quantum-motion}, we
studied numerically the role played by quantum effects on  electron
dynamics in the lowest Landau level and in the presence
of a short-range correlated random potential.

\begin{figure*}
\figs{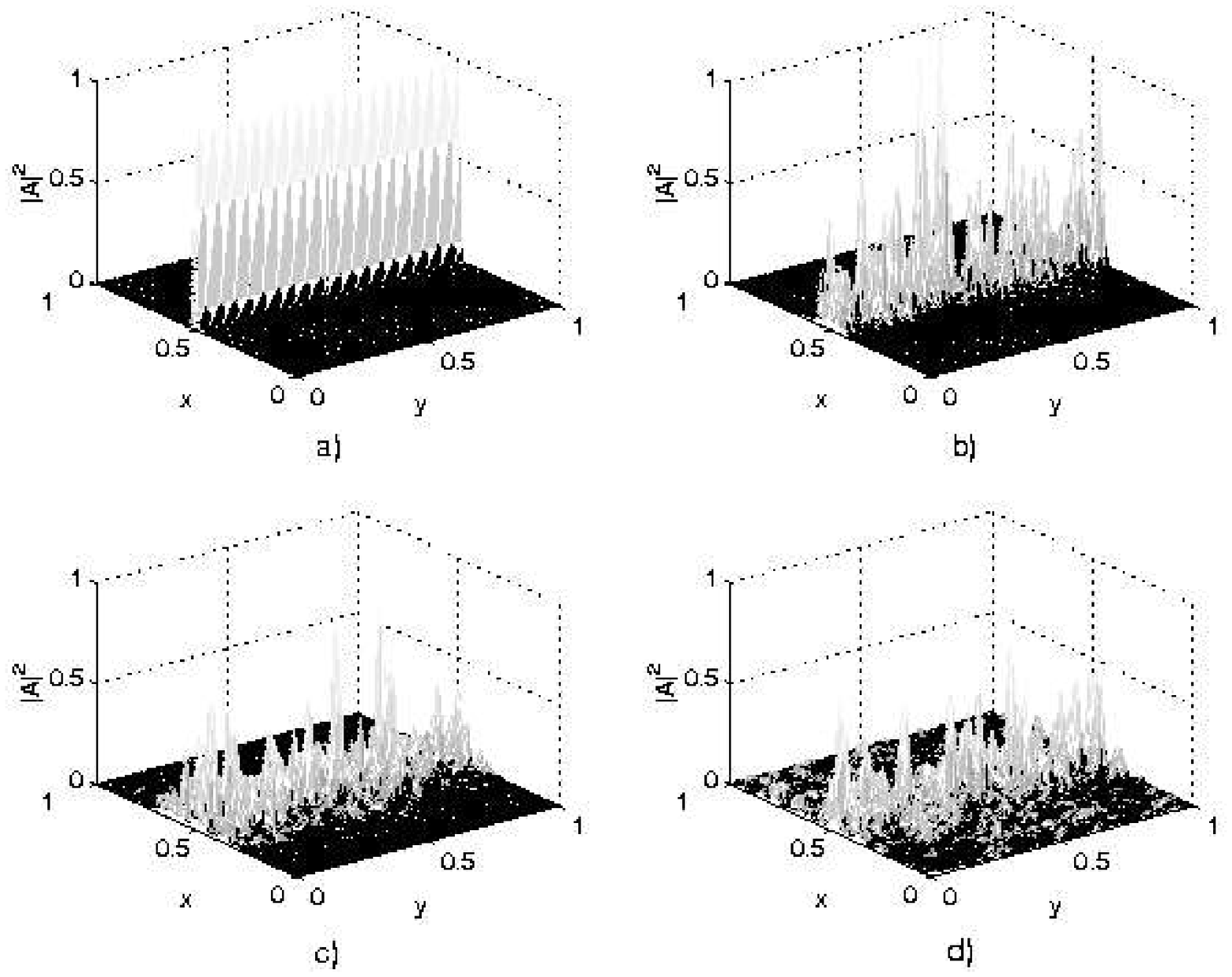}
\caption{\label{fig5:wavepack} Evolution of a wave-packet made of all
  eigenstates that is initially localized at $x = 0.5$. The squared
  amplitude of the wave-packet is plotted as a function of the real space
  coordinates $(x,y)$ at four different times.}
\end{figure*}

We use the approach proposed in Ref.~\onlinecite{ref:Gurarie} and use
the eigenstates of the Hamiltonian for an electron on a torus geometry
as the basis of states for the LLL. The corresponding wave-functions
written in the Landau gauge are:

\beqa
\psi_{\alpha}(x,y)&=&\left[\sum_{m=-\infty}^{\infty}\exp\left(2\pi(x\,+\imath y)(N\,m+\alpha)-\right.\right. \nonumber \\
&&\left. \left.\frac{(N\,m\,+\,\alpha)^2}{N}\pi \right) \right]\,e^{-\pi N x^2}
\label{eq:psialpha}
\eeqa
where $m$ takes integer values, $N$ is the number of flux quanta
through the torus, and the index $\alpha$ goes from $0$ to $N-1$,
labeling the $N$ states in the LLL. These wave-functions are periodic
functions in the interval $(x,y) \in [0,1)\times[0,1)$, and are
centered along narrow strips (of width {$\ell_c$}) around the lines $x
= \alpha/N$; an example is shown in Fig.~\ref{fig5:wavepack}a). In
Eq.~(\ref{eq:psialpha}) $(x,y)$ are dimensionless variables expressed
in units in which the magnetic length is {$\ell_c$}$ = 1/2\pi N$.

The electron density operator ${\rho}(k_1,k_2) = \mbox{exp}(2\pi i
(k_1\,x\,+\,k_2\,y))$ projected onto the LLL using this basis, is a matrix
of the form:

\beq
\bar{\rho}(k_1,k_2)\,=\,\mbox{exp}\left(-\frac{k_1^2\,+\,k_2^2}{2\,N}\pi\right)L(k_1,k_2)
\label{eq:rohat}
\eeq 
where $k_1, k_2$ are integers ($(k_1, k_2) \neq (0,0)$), and the
matrix $L(k_1,k_2)$ is given in terms of two unitary unimodular
$N$x$N$ matrices:

\beqa
L(k1,k2)\,&=&\,\epsilon^{k_1\,k_2/2}\,f^{k_1}\,h^{k_2} \nonumber \\
h\,&=&\,\left( \begin{array}{ccccc}
0&1& ... & ...0\\
0&0& 1 & ...0\\
......&...& ...& ...\\
0&...& ...& ...0 \nonumber
\end{array} \right) \\
f\,&=&\,diag(1,\epsilon,...,\epsilon^{N-1}) \nonumber
\eeqa

with $\epsilon=\mbox{exp}(\frac{2\pi i}{N})$; note that $h$ is a cyclic
permutation matrix. These matrices satisfy: $h\,f =
\epsilon\,f\,h$ and $h^N = f^N =1$. The explicit expression for  the
matrix elements of $L(k_1,k_2)$ is:

\beq
\left[L(k_1,k_2)\right]_{\alpha, \beta} = \epsilon^{k_1\,k_2/2 + k_1(\alpha-1)} \,\delta_{\alpha, \beta -k_2}\,|_{mod N}
\label{eq:Lk1k2}
\eeq

Because the basis of states is restricted to the LLL, the kinetic energy is
a constant and the projected Hamiltonian, reduces to:

\beqa \bar{H}\,&=&\,\sum_{k_1,k_2}\, V(-k_1,-k_2)\,\bar\rho(k_1,k_2)
\\ &=&\,\sum_{k_1,k_2}\,
V(-k_1,-k_2)\,\mbox{exp}\left(-\frac{k_1^2\,+\,k_2^2}{2\,N}\pi\right)L(k_1,k_2)
\nonumber
\label{eq:hamilt2}
\eeqa
where the sum runs over all integer values of $(k_1,k_2)$, and $V(0,0) = 0$
(this choice amounts to fixing the critical energy at zero).

In this basis, the Hamiltonian is a random $N$x$N$ matrix that can be
diagonalized exactly. Notice that the projection operation introduces
the exponential factor in Eq.~(\ref{eq:hamilt2}) that effectively
reduces the amplitude of the large $(k_1,k_2)$ Fourier components of
the potential (it acts as a soft large-momentum cutoff). The presence
of this factor seems to indicate that the transition is driven by the
small momenta components of the disorder potential. It is important to
remark, that even when a numerical calculation procedure fails (due to
large but finite precision) to detect the contribution coming from
momenta components higher than a maximum value for $(k_1^2+k_2^2)/2N$
(numerically, the Hamiltonian matrix becomes a banded-matrix), it is
incorrect to replace the exponential factor by a hard-cutoff, since
the commutation relations of the projected density
operator\cite{ref:Girvin1} are violated. (In numerical calculations,
this translate into having results that are extremely sensitive to the
value of the hard-cutoff chosen.)

The values for the Fourier components of the disorder potential in
Eq.~(\ref{eq:hamilt2}) are chosen randomly from a flat distribution
over the interval $[-0.5, 0.5]$ in units of $\hbar \omega_c$.

We diagonalize the Hamiltonian to obtain all $N$ eigenvalues and
eigenfunctions. Figure \ref{fig3:locals} shows examples of a localized
state at the tail of the Landau band, and a delocalized one close
to the band center, obtained with a basis of $N=1000$ states.

\befig
\figl{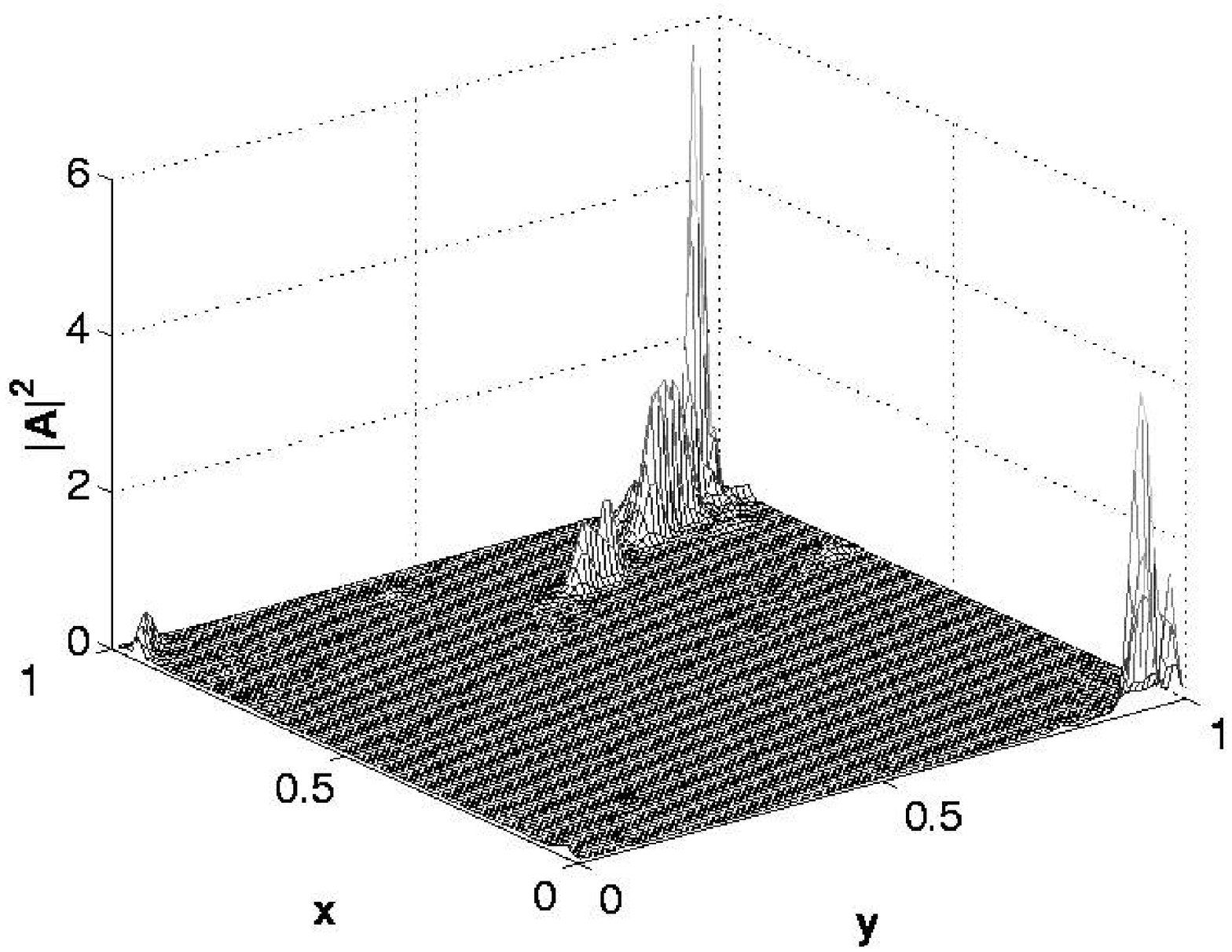}
\figl{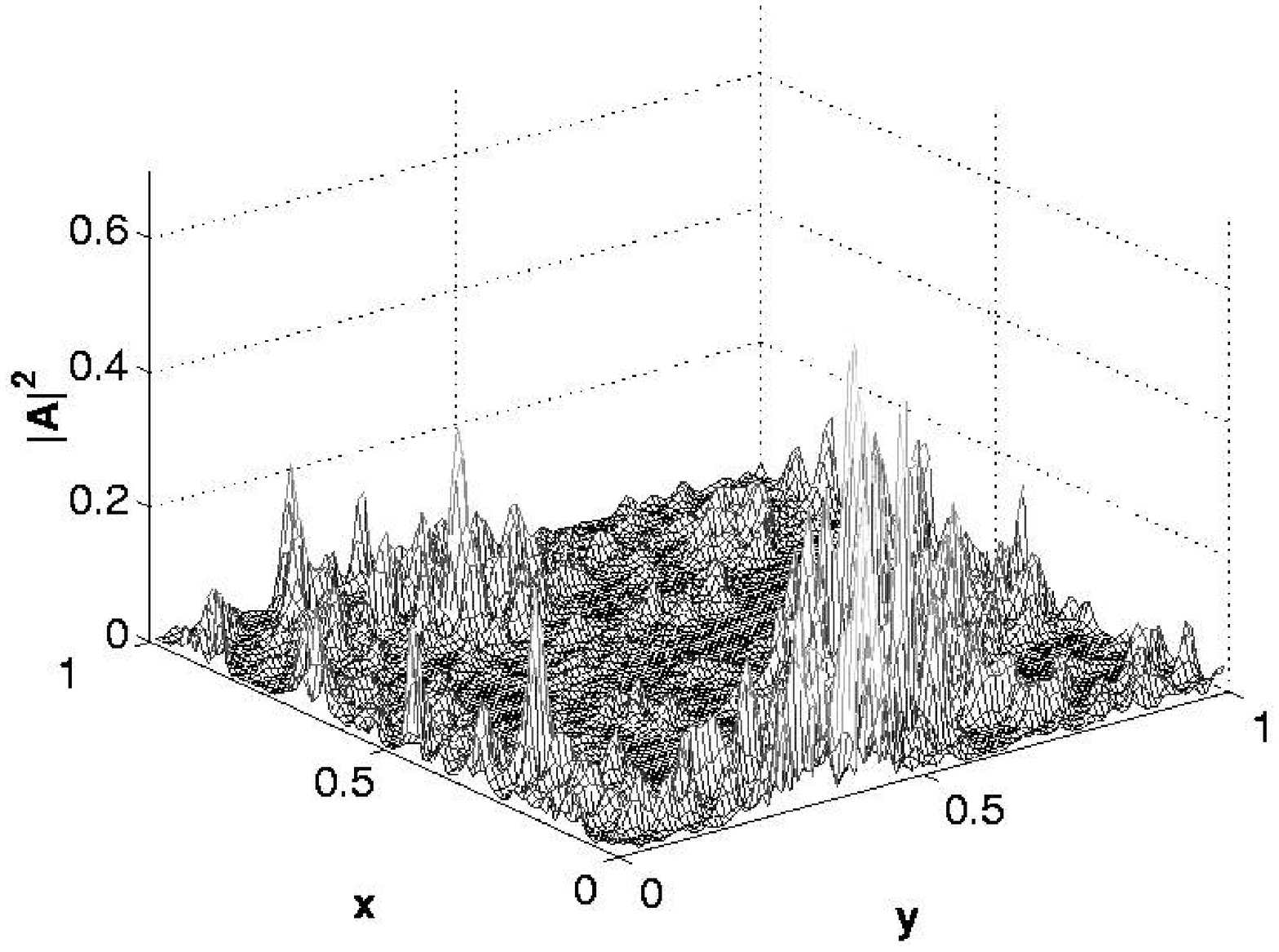}
\caption{\label{fig3:locals} Squared amplitude of a localized eigenstate
  with energy $E = 9.6$ (top), and a delocalized one with $E = 0.011$ (bottom),
  in the lowest Landau level with degeneracy $N=1000$. The
  disorder potential is short ranged.}
\efig

A localized wave-packet is constructed with all the eigenstates for a
given disorder realization and, as it evolves, its spread along the
x-direction is computed as a function of time. The wave-packets are
chosen from among the basis states, Eq.~(\ref{eq:psialpha}), which are
localized in the x-direction and completely spread out in the
$y$-direction (see Fig.~\ref{fig5:wavepack}a).  Thus, the contribution
to the total spread from the y-direction is a constant proportional to
the system size.  Fig.\ref{fig5:wavepack} shows a set of 'snapshots'
for the evolution of the wave-packet initially localized at $x = 0.5$.

The procedure is repeated for different initial positions for the
localized wave-packet and different disorder realizations. The spread
is averaged over all initial positions and disorder realizations.  A
typical result for a basis of $1000$ states and some $1000$ disorder
realizations is plotted in Fig.\ref{fig6:quansr}.

\befig
\noindent
\figl{Fig_6.eps}
\caption{\label{fig6:quansr} Spread of wave-packet as a function of
time, in the presence of short-range correlated disorder potential
with a basis of $N=1000$ states. The spread at intermediate times is
sub-diffusive. The inset shows the anomalous diffusion exponent
$\theta$ as a function of basis size $N$. The intercept gives the
infinite-system-size value of $\theta$ which leads to the quoted value
of the localization exponent $\nu$ ($\nu=1/(2-2\theta)$).}  \efig

As in the classical case, three regimes can be identified. For long
enough times ($t>10$ in Fig.\ref{fig6:quansr}) the spread of the
wave-packet reaches a constant value indicating the influence of the
finite system size.  The critical region corresponds to intermediate
times ($0.4<t<10$), and the value of $\nu_{q}$ ("q" for quantum) can
be related to the slope of the line, which is the anomalous diffusion
exponent $\theta$ (see Eq.~(\ref{eq:ThetaNu})).  As seen in the inset
to Fig.\ref{fig6:quansr}, the slope in the critical region shows a
strong dependence on the system size. To take into account these
finite size effects, we determined the slope for systems sizes ranging
from $N=200$ to $N=1500$ and compute the value for an infinite sized
system by linear extrapolation to obtain $\nu_{q} = 2.33 \pm 0.09$.

Our result compares favorably to the values measured in experiments
($\nu = 2.3 \pm 0.1$)\cite{ref:Koch} and previous numerical
simulations ($\nu = 2.35 \pm 0.05$)\cite{ref:Huckestein1}. This
provides an important check on the numerical method.

Our data shows that at very short times the slope $\theta \simeq 2$
which implies ballistic spreading of the wave-packet. In the Appendix
we present a calculation based on perturbation theory that shows that
this is a result of averaging the evolution equation over random
disorder, at short enough times.

\section{Power-law correlated disorder potential}
\label{sec:longrange}

In this section we address the central issue raised by the present
work: how a change in the properties of the probability distribution
for the disorder potential affect the nature of localization in the
lowest Landau level.  The numerical procedure outlined in previous
sections allows us to investigate this point in a straightforward
manner for both the classical and the quantum regime. Basically, it
amounts to an appropriate modification of the probability distribution
for the Fourier components for the random potential in
Eq.~(\ref{eq:hamilt2}).  Since we are interested in the effect of
power-law correlations on localization properties of the Hamiltonian
in Eq.~(\ref{eq:hamilt2}), the Fourier components are chosen
independently with a Gaussian distribution with zero mean, and
variance \beq \overline{|V({\bf k})|^2} \propto 1/|{\bf k}|^{2-\alpha}
\ .
\label{eq:VFT}
\eeq
Inverse Fourier transforming to real-space leads to power-law correlations

\beq
\overline{V({\bf r}) V({\bf 0})} \sim 1/{|{\bf r}|^{\alpha}} \ .
\label{eq:potcorr}
\eeq

The constant in Eq.~(\ref{eq:VFT}) is fixed so that the variance of
$V(r)$ is normalized to one.  Long-range correlations strongly modify
the real-space configuration of the equipotential lines as can be seen
in Fig.~\ref{fig7:s_potent} where a comparison between a short-range
and a power-law correlated potential with $\alpha = 0.5$ is given.

\befig
\figl{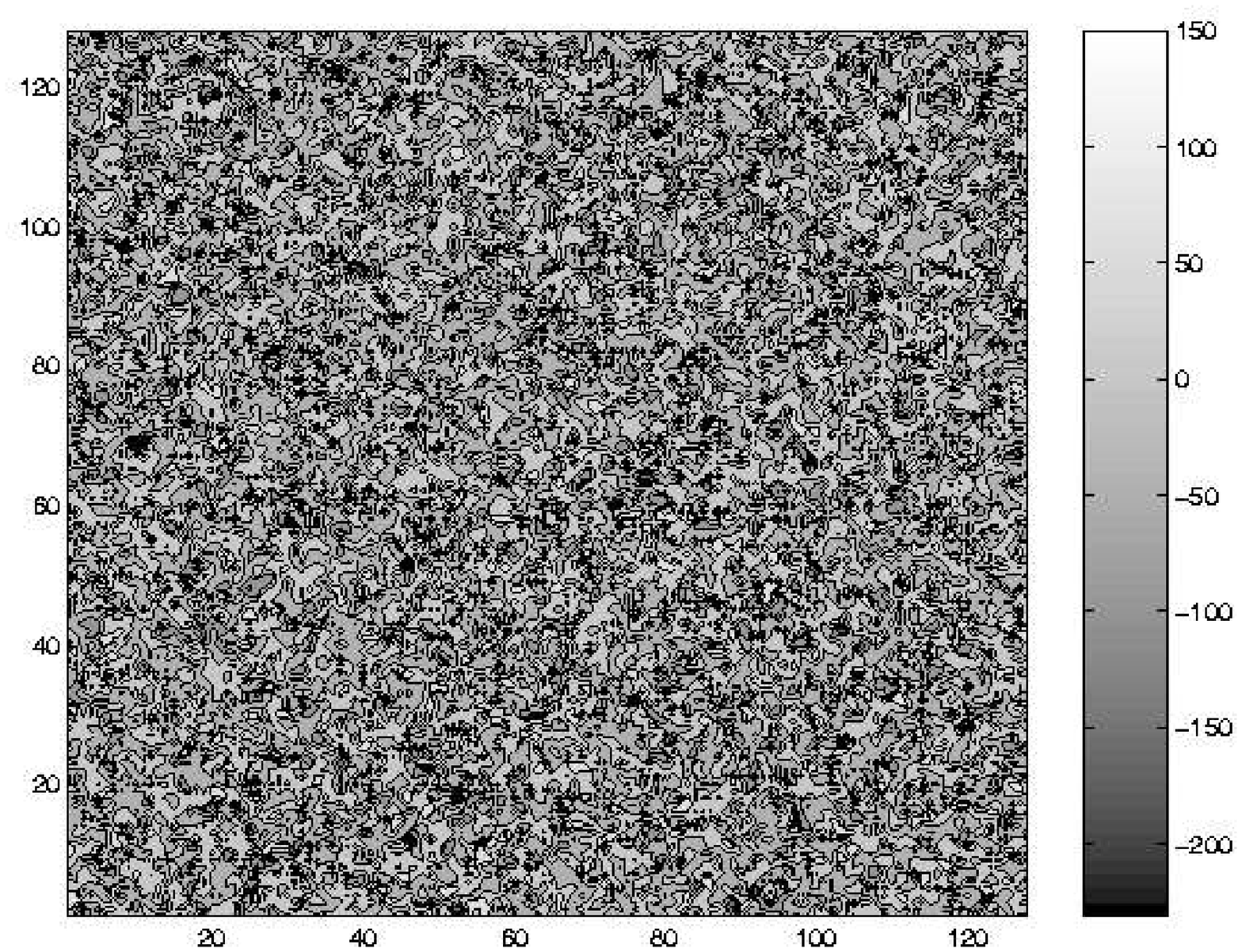}
\figl{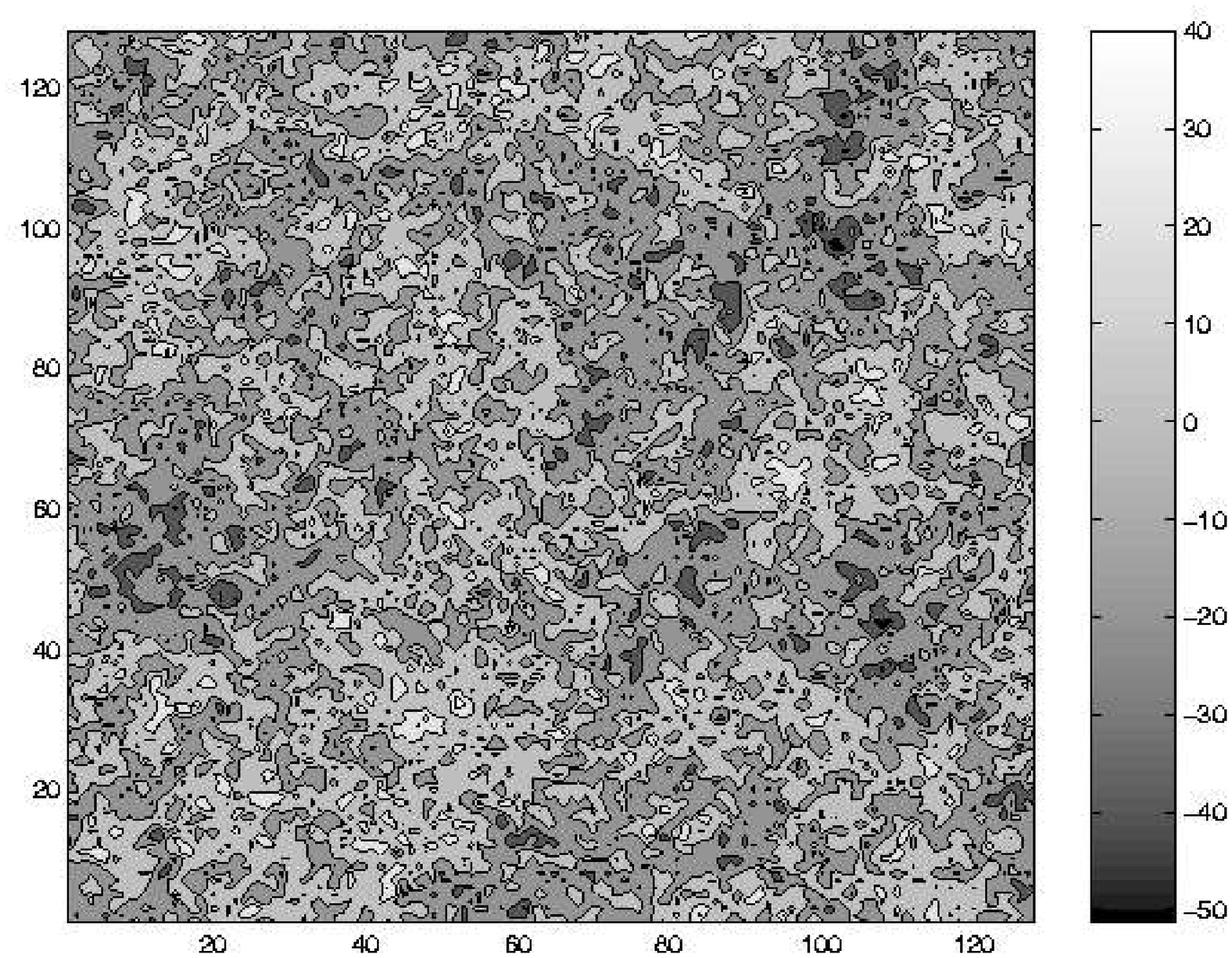}
\caption{\label{fig7:s_potent} Contour plot of a typical realization of  a short-range
correlated potential (top) and a power-law correlated one with exponent $\alpha = 0.5$.}
\efig

\subsection{Classical motion}
\label{sec:longclassical}

In the classical limit of the IQH transition, the effect of power-law
correlations in the disorder potential can be described in terms of a
purely geometric effect: they change the fractal geometry of the
equipotential contours modifying the electron's path, and
correspondingly, the anomalous diffusion law for its mean squared
displacement.

To study numerically this regime, we repeated the procedure described
in Sec.\ref{sec:shortclassical}, and computed $\overline{\langle
\Delta {\bf r}^2(t)\rangle}$ for values of $\alpha$ ranging from
$0.33$ to $1.8$. Fig.~\ref{fig9:clas_lr} shows the curves obtained for
a system of $512$ x $512$ lattice sites.  As in the case of
short-range correlations, three regimes can be identified with the
onset of each of them depending on the value of $\alpha$. From the
figure, a qualitative change in the spread of the electrons position
in time is observed. Namely, in the critical regime (typically for $30
< t < 1100$), the value of the slope becomes dependent on $\alpha$
when $\alpha \leq 1.5$. As the value of $\alpha$ decreases
(correlations increase) below a critical value of $\alpha^*_c \simeq
1.5$ ('c' is for 'classical'), the slope $\theta$ in the critical
region increases and $\nu$ becomes an increasing function of $\alpha$
tending to infinity as $\alpha$ approaches zero.

\befig
\figl{Fig_9.eps}
\caption{\label{fig9:clas_lr}Classical motion: averaged
mean square displacement as a function of time for $512\times 512$ system
size. Distances are measured in units of lattice spacing and time in
number of lattice steps. }
\efig

\subsection{Quantum motion}
\label{sec:longquantum}

In contrast to the classical regime, the effect of power-law
correlations in the disorder potential on the properties of
quantum localization, does not have a simple geometrical interpretation.
In particular, we find that changing the fractal properties of the
equipotentials does not always lead to a change in the quantum critical
exponents.

The numerical procedure introduced in Sec.~\ref{sec:shortquantum} is
easily extended to the present case allowing us to study the problem
in detail. The Fourier components of the random potential are taken
from the same distribution as in the classical case and the
corresponding quantum Hamiltonian (Eq.~(\ref{eq:hamilt2})) is
diagonalized. Then, $\overline{<\Delta x^2>}$ is computed as a
function of time for values of $\alpha$ ranging from $0.1$ to $1.9$. A
typical set of results obtained with a basis of $1000$ states is shown
in Fig.~\ref{fig10:quanlr} together with the curve obtained for a
short-range correlated potential. As in the short-range case, the
exponent $\theta$ was calculated for different systems sizes ranging
from $N=300$ to $N=1000$.

\befig
\figl{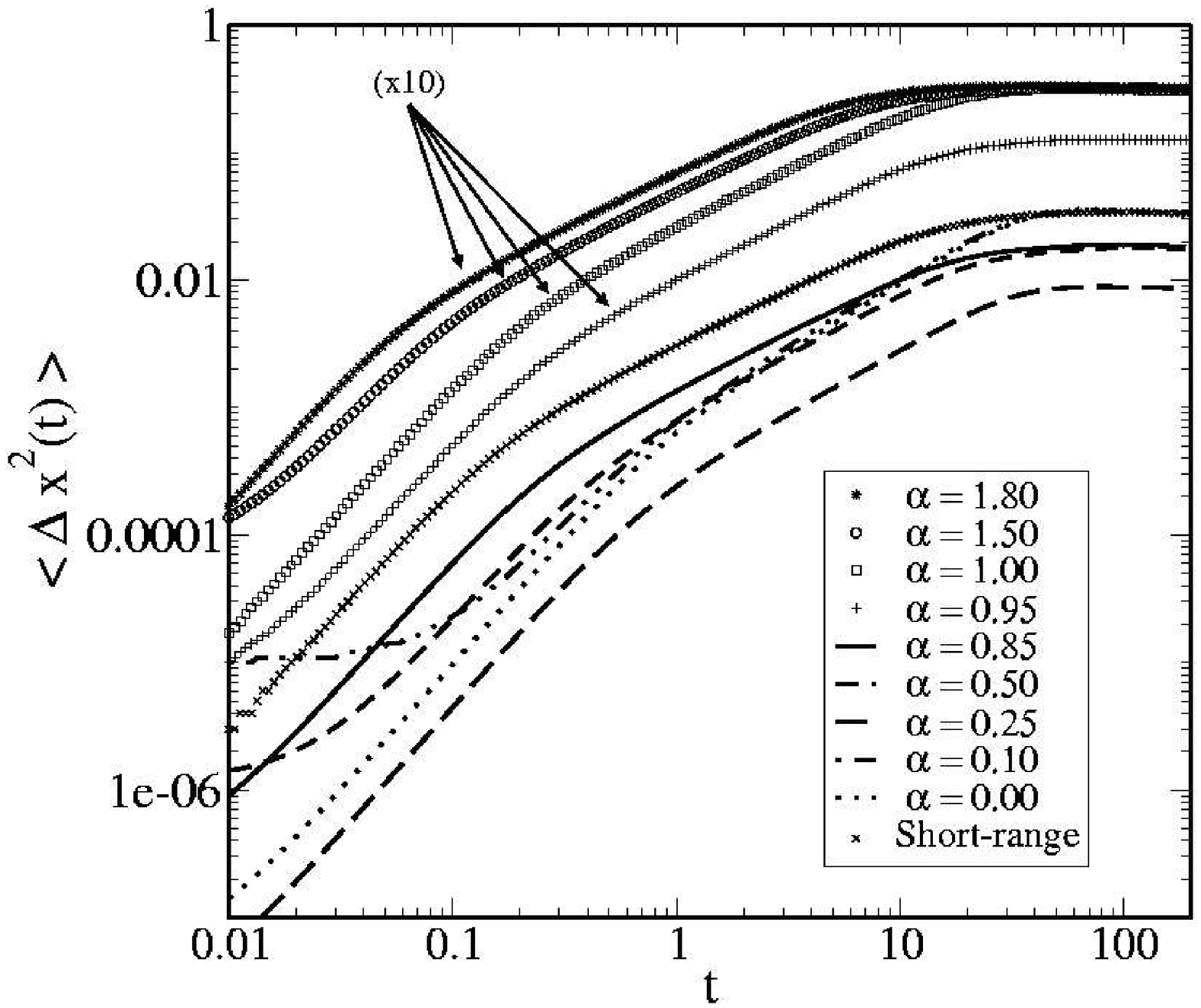}
\caption{\label{fig10:quanlr} $\overline{<\Delta x^2>}$ as a function of time
for various values of the power-law correlated disorder exponent.
Anomalous diffusion appears at intermediate times. The top four curves
have been multiplied by a factor of 10 for clarity.}
\efig

Like in the classical case, a qualitative difference in the slopes of
the critical regime develops as $\alpha$ changes. A careful comparison
among the different curves shows that the value of the slope in the
critical region starts to increase below a critical value of
$\alpha^*_{q} \simeq 0.85$. A detailed quantitative analysis of these
results together with a comparison to the classical case is the
subject of the following section.

\section{Extended Harris criterion}
\label{sec:HarrisCrit}

The effect of a random disorder potential on the properties of
continuous phase transitions in classical systems is summarized by the
Harris criterion \cite{ref:Harris}. Basically, the criterion states
that critical exponents of the disordered and the corresponding clean system
remain equal as long as the value for the correlation length exponent $\nu$
satisfies: $d \nu - 2 \geq 0$, where $d$ is the dimensionality of the
classical system under consideration. The criterion is derived by
requiring that the fluctuations of the thermodynamic quantity which
couples to the disorder, within a volume set by the correlation
length $\xi$, do not grow faster than its average value, as the
transition is approached and $\xi \to \infty$. It was proved rigorously and
generalized to a wider class of systems if an appropriate
definition for the correlation length is adopted \cite{ref:Chayes}.
An extension of the criterion to power-law correlated disorder was
proposed by Weinrib and Halperin \cite{ref:Weinrib1}.

In the case of two-dimensional percolation with power-law correlations
in the occupation probabilities, with the power-law exponent $\alpha$,
the extended Harris criterion states:

\beqa
\alpha &>& {\alpha^*_{c}=3/2} \;\;\;\;\;\;\;\; \nu_{c}= \frac{4}{3} \nonumber \\
\alpha &<& {\alpha^*_{c}=3/2}  \;\;\;\;\;\;\;\;\nu_{c}= \frac{2}{\alpha} \ .
\label{eq:clas_harris}
\eeqa

As a numerical check on Eq.~(\ref{eq:clas_harris}), we computed the
values of the anomalous diffusion exponent $\theta$ for classical
electron motion and plotted them as a function of $\alpha$ in
Fig.~\ref{fig11:clas_nu_vsal}.  The values of $\theta$ were obtained
from fitting the data in Fig.~\ref{fig9:clas_lr}, in the critical
region, to a line using a least-squares method; $\theta$ is the slope
of the line.  We see that the data is well described by the
theoretical prediction: above $\alpha^*_{c} = 3/2$ the anomalous
diffusion constant $\theta=1-1/2\nu_c=5/8$ is constant, while below
this value it varies as $\theta=1-\alpha/4$.  Similar results were
obtained previously by Prakash {\em et al.} \cite{ref:Prakash}.

\befig
\figl{Fig_11.eps}
\caption{\label{fig11:clas_nu_vsal} Classical regime: exponent
$\theta$ as a function of power-law correlated disorder exponent
$\alpha$. The horizontal line represents the theoretical value of
$\theta = 0.625$ corresponding to the value of $\nu_{c}= 4/3$
obtained from 'classical' percolation theory. The linear dependence
of $\theta=1-\alpha/4$ is the prediction of the extended Harris
criterion, Eq.~(\ref{eq:clas_harris})}.
\efig

Analogous results hold true for the quantum case.  In
Fig.~\ref{fig12:quant_nu_vsal} we plot the anomalous diffusion
exponent $\theta$, which describes the spreading of an initially
localized wave-packet in the lowest Landau level, as a function of
$\alpha$.  The exponent $\theta$ was computed using the data in
Fig.~\ref{fig10:quanlr} following the same procedure as in the
classical case. The values plotted are the extrapolation corresponding
to an infinite system size.  We see that the numerical results
presented in Fig.~\ref{fig12:quant_nu_vsal} are in good agreement with
the extended Harris criterion which now reads:

\beqa
\alpha &>& {\alpha^*_{q}\approx 0.85} \;\;\;\;\;\;\;\; \nu_{q} \approx 2.33  \nonumber \\
\alpha &<& {\alpha^*_{q}\approx 0.85}  \;\;\;\;\;\;\;\;\nu_{q}= \frac{2}{\alpha} \ .
\label{eq:quant_harris}
\eeqa

Support for the validity of the extended Harris criterion for quantum
critical points was also provided by studies of random quantum magnets
\cite{ref:Rieger}.

\befig
\figl{Fig_12.eps}
\caption{\label{fig12:quant_nu_vsal} Quantum regime: anomalous
diffusion exponent $\theta = 1- 1/2\nu_q$ as a function of the
exponent $\alpha$, which determines the power-law correlations of the
disorder. The lines are predictions of the extended Harris criterion
(Eq.~(\ref{eq:quant_harris})) using the value $\nu_q=2.33$ for the
quantum localization length exponent, in the case of short-range
correlated disorder.}
\efig

In the case of quantum percolation (i.e., localization in the lowest
Landau level) the extended Harris criterion can be argued for in close
analogy to the classical case \cite{ref:Weinrib1}. Namely, consider
the average size of the fluctuations of the electron's energy in an
area set by the localization length $\xi$. Like the electron's energy,
it is also determined by the random disorder potential, and can be
approximated as: 

\beqa 
\overline{E^2} & \approx & \overline{\left(
\frac{1}{\xi^2} \int_{\xi^2} d^2x V({\bf x}) \right )^2} \nonumber \\ 
& \sim & \frac{1}{\xi} \int_{\xi^2} d^2x |{\bf x}|^{-\alpha}
\label{eq:harris1}
\eeqa
where in deriving the second equation we have
made use of $<V({\bf x}) V({\bf x}')> \sim 1/|{\bf x}-{\bf x}'|^{\alpha}$.

For the quantum critical point associated with localization in the
lowest Landau level to be stable to the introduction of power-law
correlations in the random potential, the fluctuations in the
electron's energy should be small when compared to its energy
$E$. Using Eq.~(\ref{eq:harris1}), and the scaling relation $E\sim
\xi^{-1/\nu}$ we arrive at the estimates

\beq \frac{\overline{E^2}}{E^2} \sim \left\{
\begin{array}{r@{\quad:\quad}l}
E^{2\nu - 2}  & \alpha > 2 \\
 E^{2\nu - 2} ln(E)^{-\nu} & \alpha = 2 \\
 E^{\alpha \nu -2} & \alpha < 2  \ . \nonumber
\end{array} \right.
\eeq

The quantum percolation critical point is expected to be unaffected by
the introduction of power-law correlations if $\overline{E^2}/E^2 \to
0$ when $E\to 0$, or, equivalently, when $\xi\to \infty$.  Since
$\nu>1$ this is always the case when $\alpha \geq 2$.

However for $\alpha< 2$ there are two regimes to consider.  When $2 >
\alpha > 2/\nu$, $\overline{E^2}/E^2 \to 0$ and the localization
critical point due to short-ranged potential disorder is again stable
to introduction of power-law correlations in the random potential. In
contrast, for values of $\alpha < 2/\nu$ a power-law correlated
potential produces large fluctuations in the energy thus destabilizing
the critical point.

The usual expectation is that this relevant perturbation leads to a
new quantum critical point characterized by a new value of $\nu$. This
is confirmed by our numerical results.

An important consequence of the present result regards the validity of
the expression put forward in Ref.~\onlinecite{ref:Milnikov}, that
relates the values of $\nu_{q}$ and $\nu_{c}$ as follows:

\beq
\nu_{q} = \nu_{c} + 1
\label{eq:nuq_plus_nc}
\eeq

According to the analysis presented above, a quantum system with a
long-range correlated potential has the same critical exponent as one
with short-range correlations as long as $\alpha \geq 0.85$. However,
a classical system with potential correlations with a value for
$\alpha$ in the region $0.85 < \alpha < 1.5 $, has a continuously
varying critical exponent $\nu$ according to
Eq.~(\ref{eq:clas_harris}). For instance, a value of $\alpha = 1$ gives
$\nu_{q} = 2.33$ while $\nu_{c} = 2.0$. This analysis suggests that
the relation given by Eq.~(\ref{eq:nuq_plus_nc}) (whatever the argument
used to support it) is inconsistent with the Harris criterion.

\section{Competing disorders}
\label{sec:short-long-correl}

In this section we undertake a numerical study of the effect of
competing disorders on localization in the LLL. We investigate first
the classical regime, which is in the universality class of correlated
percolation. Analytical calculations \cite{ref:Weinrib2} have shown
that in these systems the classical critical point generated by a
long-range correlated potential is stable against perturbations
produced by a short-range correlated potential. As shown below, our
numerical results in this regime, provide support for this picture.

In the quantum case we investigate the effect of competing power-law
correlated and short-range correlated disorders on localization.
Numerical studies of this particular case are relevant for experiments
attempting to measure properties of the correlated quantum percolation
critical point. Namely, since short-range correlated potentials are
always present in experimental setups, it is important to understand
quantitatively how their presence affects the values of the new
localization length exponents described in the previous sections.

To proceed with the appropriate numerical model we first establish the
following conventions.  The total disorder potential is defined as a
weighted sum of two potentials: one with short-range correlations
$V_s({\bf r})$ and one with long-range correlations $V_l({\bf
r})$. The units chosen are such that the mean values of short- and
long-range correlated potentials are set equal to zero while their
variances are normalized to one. In this way, the maximum amplitudes
of both potentials are comparable.  In order to vary the relative
strength of each potential we introduce a parameter $\kappa$ with
values in the range $0 \leq \kappa \leq 1 $ such that the total
potential is:

\beq
V_{eff}({\bf r}) = \kappa \; V_s({\bf r}) +
(1-\kappa) \; V_l({\bf r}) \ .
\eeq

Notice that when $\kappa = 0$ the total potential has power-law
correlations only, while the short-ranged part is increasingly
dominant as the value of $\kappa$ increases toward 1. This definition
ensures that the variance of the total potential $V_{eff}$ remains
fixed as $\kappa$ changes and $\langle V_{eff}^2 ({\bf r}) \rangle =
1$.

The numerics were performed with short-range correlated potentials
constructed from random Fourier components $V({\bf k})$ sampled from a
uniform distribution, and from a Gaussian distribution with variance
$\sim |{\bf k}|^{\alpha-2}$ with $\alpha>\alpha^*$.  In real space
this leads to delta-correlations and power-law correlations with
exponent $\alpha$, respectively.  Long-range correlated potentials
were generated with power-law correlations with exponent
$\beta<\alpha^*$, for which the extended Harris criterion predicts a
new localization length exponent $\nu=2/\beta$. The properties of the
total potential and its components is summarized in Table~\ref{table}.

\begin{table}
\caption{Statistical properties of the short-range and long-range correlated
  potentials used in section \ref{sec:short-long-correl}.
\label{table}}
\begin{ruledtabular}
\begin{tabular}{ccl}
$V_{eff}({\bf r})$ &=& $\kappa \; V_s({\bf r}) +
(1-\kappa) \; V_l({\bf r})$ \nonumber \\
$\langle V_s({\bf r}) \rangle = \langle V_l({\bf r}) \rangle$ &=& 0 \nonumber  \\
$\langle V_s^2({\bf r}) \rangle = \langle V_l^2({\bf r}) \rangle$ &=& 1 \nonumber  \\
$\langle V_s({\bf r}) V_s({\bf r'}) \rangle$ & $\propto$ &
$|{\bf r}-{\bf r'}|^{-\alpha}$ , $\;\;\; \delta({\bf r} - {\bf r'})$ \nonumber  \\
$\langle V_l({\bf r}) V_l({\bf r'}) \rangle$ & $\propto$ &
$|{\bf r}-{\bf r'}|^{-\beta}$  \nonumber  \\
\end{tabular}
\end{ruledtabular}
\end{table}

A note on the evaluation of errors: as it will be shown in the
following sections, the effect of tuning $\kappa \neq 0 $ depends on
system size for both classical and quantum regimes. Hence, it was not
possible to carry out the finite-size study similar to the one done
previously, where there is only one type of disorder present. As a
consequence, the errors have been estimated from the largest
systematic error involved in the procedure, which appears in the
determination of the slope of the critical region.

\subsection{Classical Regime}
\label{sec:clas-mixed}

Computations were carried out for system sizes $N = 256, 512$ and
$N=1024$.  The values of the anomalous diffusion exponent $\theta$
were extracted from the critical region of the spread of the classical
position of the electron, as in section
\ref{sec:longclassical}. Values of $\theta$ were obtained for a range
of values of the parameter $\kappa$.  Fig.~\ref{fig:clas-long-short}
shows result that correspond to a combination of a power-law
correlated potential with exponent $\beta = 0.5$ (for the long-range
part) and $\alpha = 1.8$ (for the short-range piece).

\befig
\figl{Fig_13.eps}
\caption{\label{fig:clas-long-short} Classical regime: slope
$\theta$ as a function of parameter $\kappa$. $\kappa = 0 (1)$ indicates
a pure long-range (short-range) correlated potential.}
\efig

The figure shows that, as the system size increases the value of
$\theta$ (and hence of $\nu$) remains closer to the pure long-range
value up to fairly large values of $\kappa$ (roughly
$55\%-60\%$). The increasing sharpness of the crossover with
increasing system sizes suggests a very sharp crossover in
the thermodynamic limit. In the language of renormalization group,
this is an indication of the stability of the long-range
critical point when perturbed with a short-range correlated potential.
Notice, as it was pointed out above, that this crossover is strongly
dependent on the system size, rendering useless the finite-size
scaling analysis used previously.  An interesting effect is observed
in the raw data that points to the way the crossovers occur in
finite size systems.  Between the initial crossover from diffusive to
sub-diffusive motion and the final crossover from sub-diffusive motion to
saturation, there are two clearly distinguishable regimes. At the
earlier times, the slope is determined by the short-range correlated
potential. Only at latest times (and before saturation effects
dominate), a new slope appears that corresponds to the value of
$\theta$ determined by the exponent of the power-law (long-range)
correlated potential.

\subsection{Quantum Regime}
\label{sec:quant-mixed}

In contrast to the classical regime, up to date, there is no
experimental or theoretical work to our knowledge that attempted to
investigate the influence of mixed long-range and short-range
correlated disorder potentials on the integer quantum Hall
transitions. The numerical strategy used in section
\ref{sec:longquantum}, provides us with a framework to study this
situation and allows us to examine the stability of newly found
quantum critical points.

Computations for quantum systems were carried out with a basis of
$N=500, 700$ and $N=1000$ states. The values of the slope $\theta$
were extracted from the critical region and plotted as a function of
the parameter $\kappa$, as in the classical regime. The results shown
in Fig.~\ref{fig:quant-long-short} were obtained with power-law
correlated potentials $ V_s $ and $ V_l $ with exponents $\alpha =
1.8$ and $\beta = 0.5$, respectively.

\befig
\figl{Fig_14.eps}
\caption{\label{fig:quant-long-short} Quantum regime: slope
$\theta$ as a function of parameter $\kappa$. $\kappa = 0$ indicates
a purely long-range correlated potential, $\kappa = 1$ corresponds to
a purely short-range correlated potential.}
\efig

As the figure shows, data for the quantum regime is noisier than its
classical counterpart. Finite size effects are also more prominent as
can clearly be observed in the variation of the values of the slope
$\theta$. However, the trend observed is in agreement with the
expected behavior in the thermodynamic limit. It suggests that the new
quantum correlated critical points are indeed stable against the
perturbation introduced by a short-range correlated potential. This
opens up the prospect of measuring the new localization length
exponents $\nu_q$ in experiments in which random disorder with
power-law correlations is introduced. The idea would be to engineer a
random potential with desired properties which would then compete with
the short-range disorder due to the presence of impurities in the
semiconductor heterostructure. An example of such a system was
described in Ref.~\onlinecite{ref:Zielinski} where a thin magnetic
film was placed in close proximity to the two-dimensional electron gas
resulting in a random magnetic field. By controlling the height
fluctuations of the film one can in principle engineer a random
potential with desired correlations.

\section{Conclusions}
\label{sec:conclusions}

We have analyzed the effect of spatial correlations of the disorder
potential on the localization-delocalization transition in the integer
quantum Hall system. The main purpose of the analysis was to
understand separately the role played by quantum effects and disorder
on the critical properties of the transition by comparing classical
and quantum localization.

The analysis of electron localization in the classical case reveals
that the localization length exponent is completely determined by the
statistical properties of the level lines of the random
potential. Using the tools of percolation theory, we were able to
implement well established numerical methods to perform a detailed
check of the validity of the Harris criterion. We found numerical
support for the extended version of the criterion, confirming previous
theoretical arguments \cite{ref:Weinrib2}.

The quantum regime was analyzed by studying the real time
density-density propagator as proposed in
Ref.\onlinecite{ref:Sinova}. This method, originally introduced as an
alternative way to calculate the critical exponent $\nu$ for a
short-range correlated potential, proved to be an excellent testing
ground to analyze the role of disorder. We obtained (following a
rather simplistic finite size analysis) a value for $\nu = 2.33 \pm
0.09$ remarkably close to experimentally measured values and
numerically calculated ones. We also found that when the decay of
power-law correlations of the random potential is slow enough this can
destabilize the quantum critical point and lead to new critical
behavior, as predicted by the extended Harris criterion.

Comparison between classical and quantum regimes with long-range
correlated disorder potentials, indicate that the effect of disorder
correlations are qualitatively similar for quantum and classical
systems. The main difference between these two cases seems to be the
critical value of $\alpha$ (the exponent determining the long-range
correlations) below which the value of $\nu$ is changed. The value of
$\alpha^*_{q} \simeq 0.85$ is smaller compared to its classical
counterpart $\alpha^*_{c} = 1.5$.  This suggests that quantum
fluctuations can be thought of as effectively smearing out the
correlations in the random potential thus shifting $\alpha^*$.

An immediate consequence of these results is that in the quantum case,
there is no direct connection between the statistical properties of
equipotential lines and the localization length
exponent. Specifically, we were able to show that the fractal geometry
of the equipotential lines can be continuously varied while the
quantum localization length exponent does not not change.

The detailed numerical study of the variation of the critical exponent
$\nu$ with the power-law exponent $\alpha$, led us to propose a
precise statement for the quantum version of the extended Harris
criterion. These numerical results are supported by a scaling
argument, in close analogy with the classical case.

In order to address the experimentally relevant situation of samples
with disorder potentials of different origins and likely, different
correlations, we analyzed the effect of competing correlations in
disorder potentials on the IQHT.  In the semi-classical limit, our
numerical results give support to previous theoretical arguments for
the stability of the classical correlated-percolation critical points.
In the quantum regime we observed that, while the value of the
critical exponent $\nu$ varies as the short-range correlated potential
is introduced, the trend is similar to the one observed in the
classical regime. As a consequence, our results show no qualitative
difference between classical and quantum regimes regarding the
stability of these critical points. More importantly, they point to
the possibility of seeing effects of the correlated quantum
percolation critical points in experiments on two-dimensional electron
gases in semiconductor heterostructures where short-range disorder due
to impurities is unavoidable.

It is a pleasure to acknowledge useful conversations with B. Halperin,
J. Sinova, V. Gurarie, S. Boldyrev, J. Moore, H. Castillo, S. Simon
and G. Murthy.  JK is supported by the NSF under grant number
DMR-9984471, and is a Cottrell Scholar of Research Corporation.

\appendix*
\section{}

In this appendix we present a calculation based on perturbation theory
that shows that at short times the wave-packet has a ballistic spread,
which is what was observed numerically.  If $|\psi(t) \rangle$ is the
wave-function of the wave-packet at time $t$, the spread is given by:

\beq
\langle \psi(t) | \Delta x^2 | \psi(t)
\rangle =  \langle \psi(t) | x^2 | \psi(t) \rangle -  \langle \psi(t)
| x | \psi(t) \rangle ^2
\eeq

Here we analyze the first term in detail, while the second one can be
dealt with in a similar way.

In the Heisenberg representation:

\beq
\langle \psi(t) | x^2 | \psi(t) \rangle = \langle \psi(0) | x^2(t) |
\psi(0) \rangle
\eeq
where
\beq
x^2(t) = e^{i H t }\; x^2(0)\; e^{-i H t} \ .
\eeq

Using the well known operator identity:

\begin{eqnarray}
&&\langle \psi(t) | x^2 | \psi(t) \rangle = \langle \psi(0) | x^2(0) | \psi(0) \rangle \\ \nonumber
&+& i\;t\; \langle \psi(0) |[H; x^2(0)] |\psi(0) \rangle \\ \nonumber
&+& \frac{(i\;t)2}{2!}\;\langle \psi(0) |[H;[H; x^2(0)]] |\psi(0) \rangle
+ \mathcal{O}(t^3) .
\end{eqnarray}

The probability distribution of energies is determined by the
probability distribution of the disorder potential, which is chosen to
be symmetric around the value $V=0$. Since the wave-packet contains all
the eigenstates, the averaging over disorder (average over energies)
eliminates the linear term in the expansion above. Thus, the first
non-vanishing contribution at short times is given by the quadratic
term which corresponds to a ballistic spread.

\bibliography{references}

\end{document}